\documentclass[fleqn,10pt]{wlscirep}
\usepackage[utf8]{inputenc}
\usepackage[T1]{fontenc}

\title{Seasonality and Light Phase-Resetting in the Mammalian Circadian Rhythm}

\author[1,*]{Kevin M. Hannay}
\author[2,4,+]{Daniel B. Forger}
\author[2,3,+]{Victoria Booth}
\affil[1]{Schreiner University, Department of Mathematics, Kerrville, 78028, USA}
\affil[2]{University of Michigan, Department of Mathematics , Ann Arbor, 48109, USA}
\affil[3]{University of Michigan, Department of Anesthesiology , Ann Arbor, 48109, USA}
\affil[4]{University of Michigan, Department of Computational Medicine and Bioinformatics, Ann Arbor, 48109, USA}

\affil[*]{khannay@schreiner.edu}

\affil[+]{these authors contributed equally to this work}

\begin{abstract}
We study the impact of light on the mammalian circadian system using the theory of phase response curves. Using a recently developed ansatz we derive a low-dimensional macroscopic model for the core circadian clock in mammals. Significantly, the variables and parameters in our model have physiological interpretations and may be compared with experimental results. We focus on the effect of four key factors which help shape the mammalian phase response to light: heterogeneity in the population of oscillators, the structure of the typical light phase response curve, the fraction of oscillators which receive direct light input and changes in the coupling strengths associated with seasonal day-lengths. We find these factors can explain several experimental results and provide insight into the processing of light information in the mammalian circadian system. In particular, we find that the sensitivity of the circadian system to light may be modulated by changes in the relative coupling forces between the light sensing and non-sensing populations. Finally, we show how seasonal day-length, after-effects to light entrainment and seasonal variations in light sensitivity in the mammalian circadian clock are interrelated. 
\end{abstract}
\begin{document}

\flushbottom
\maketitle

\thispagestyle{empty}

\section*{Introduction}
Daily or circadian cycles in behavior and metabolism can be observed for virtually all forms of life. The utility of circadian rhythms relies on the proper timing of these cycles relative to external environmental oscillations. Thus, a defining property of circadian rhythms is their ability to be entrained to external time cues or zeitgebers.  The principal zeitgeber for the mammalian circadian clock is light \cite{Pittendrigh1976a}. Therefore, a crucial component to understanding mammalian circadian rhythms is an improved understanding of the impact of light on the circadian cycle. A first step in this endeavor is understanding the response of the circadian circuit to a brief light pulse. 

The theory of phase response curves (PRC) provides a natural language for studying the effects of external stimuli on endogenous rhythms with a rich history of application to circadian biology \cite{Johnson1999, Winfree1980}.  Phase response curves characterize the phase shift induced by the application of the stimulus at different phases of the oscillation. For instance, the amplitude of the PRC gives the entrainment range of the system to a weak resetting signal and the zeros of the PRC specify the entrainment angle \cite{ArkadyPikovskyMichaelRosenblum2003}. 

Phase response curve theory figured prominently in early investigations of circadian rhythms, where organisms were exposed to a sensory stimulus at a sampling of points across the daily cycle and phase shifts were measured relative to to some behavioral or physiological marker \cite{Daan2001, Winfree1980}. However, the discovery of the location of the master circadian clock in a small region of the hypothalamus known as the suprachaismatic nucleus (SCN) provided a neurological basis for circadian rhythms in mammals \cite{Moore1972, Stephan1972}. The SCN was found to contain thousands of coupled clock neurons which each contain a biochemical oscillator with a period of approximately twenty-four hours \cite{Liu1997a}.  Daily activity cycles are driven by this large ensemble of coupled oscillators acting collectively to produce a reliable circadian oscillation. 

Thus, a light stimulus applied to the mammalian circadian rhythm does not act by shifting a single limit cycle oscillator, but rather acts by shifting the oscillations of individual clock neurons which in turn induce a shift in the collective rhythms produced by the ensemble.  The recognition of this distinction helped motivate the development of the theory of collective phase response curves which describe the collective phase shift of an entire population of coupled oscillators subjected to a stimulus \cite{Kawamura2008, Ko2009, Kori2009, Levnajic2010}. Collective phase resetting is especially important for the mammalian circadian response to light, because only a fraction of the clock cells are phase shifted in response to the stimulus  \cite{Meijer2003}. This may induce non-trivial transient dynamics on the system following a light perturbation and forms a major focus of this work \cite{Nagano2003}.

In general, the phase-shifting behavior of a coupled ensemble of oscillators differs from the behavior of a single autonomous oscillator. In this work we study the transformation between the response of a single circadian cell to a light-pulse (microscopic PRC) and the collective phase response described by the shift in the mean-phase of the population of oscillators. A growing literature on collective phase resetting has revealed that coupling, oscillator heterogeneity and network structure can all lead to significant differences between the microscopic and collective PRCs in networks of coupled oscillators \cite{Levnajic2010, Kori2009, Ko2009, Kawamura2008}. Within the circadian literature, examinations of collective phase resetting have led to the formation of a rule of thumb that increasing phase dispersion in the oscillator population leads to a monotonic increase in the amplitude of the collective phase response \cite{Pittendrigh1991, Abraham2010} and thus the entrainment range of the collective oscillator \cite{Winfree1980}. 

In addition to the role of light input in ensuring the circadian clock is synchronized to the outside environment, the SCN is also responsible for storing seasonal day-length information \cite{Coomans2015, Meijer2007, Sumova2004}. The ability of seasonal day-lengths to alter the core circadian clock was established in early circadian studies, where it was noticed that entrainment of mammals to long/short day-lengths caused lasting changes in the endogenous circadian period when organisms were transfered to a dark environment \cite{Pittendrigh1976f}. These effects are known as seasonal after-effects and have been described for many mammal species \cite{Daan2001, Pittendrigh1976f}.  

Recently, significant progress has been made in characterizing the physiological changes in the SCN underlying seasonal day-lengths changes \cite{VDL2009, Myung2012, Myung2015, Evans2013, Buijink2016}. The physiological changes in the SCN which encode the seasons have been shown to affect the phase response to brief light pulses. When organisms are entrained to long (summer) days the phase shift caused by a brief light pulse is seen to decrease \cite{Ramkisoensing2014,VDL2009}. This seemingly contradicts the rule of thumb for collective phase resetting, because experimental evidence has also shown that phase dispersion in the SCN increases in longer day-lengths \cite{VDL2009}. A primary goal of this work is to provide a unified theory of collective phase resetting to light in mammals, consistent with seasonal changes in SCN physiology.

In order to study phase resetting to light we make use of the recently developed $m^2$ ansatz to derive a three-dimensional model for the the core circadian clock \cite{Hannay2018}. Significantly, the $m^2$ ansatz is supported by experimental evidence and the resulting model gives variables and parameters which may be interpreted physiologically in the SCN \cite{Hannay2018, Lu2016}. Our work utilizes the framework for studying collective phase resetting developed in Levnajic and Pikovsky \cite{Levnajic2010}, and we apply this theory specifically to light induced phase resetting in mammals. We extend the results in Hannay et al \cite{Hannay2015} to consider oscillator networks where only a fraction of the population receives the phase-shifting stimulus. Additionally, we develop a perturbation technique which can be applied generally to characterize the effects of coupling on collective phase resetting.

The principal biological impacts of our work are three fold.  First, we provide a general theory for the effect of the collective amplitude on the phase-shifting capacity of the circadian clock. Our analysis reveals that the rule of thumb that lower amplitude rhythms with larger phase dispersion give larger phase shifts in response to a stimulus is incomplete, and more detailed analysis is required for many real-world phase response curves. Secondly, our analysis reveals the reduction in light-shifting capacity observed for organisms entrained to long day-lengths may be explained by an adjustment of the coupling strengths between clock neurons as a function of the seasonal day-length. Finally, we find this adjustment of coupling strengths is consistent with current theories for seasonal day-length encoding and is required to explain seasonal after-effects to light entrainment in mammals.


\section*{Results}

\subsection*{Formulation of the Model}
\label{SEC:model}

\subsubsection*{Circadian Model}
The suprachaismatic nucleus (SCN) is a collection of about $10,000$ neurons which form the core circadian pacemaker in mammals.  Individual neurons in the SCN contain a biochemical feedback loop which cycles with a period of approximately $24$ hours. While the SCN produces a variety of spatiotemporal activity patterns it can be functionally and physiologically broken into the ventral (core) and dorsal (shell) populations \cite{Foley2011}. 

\begin{figure}
\centering
\includegraphics[]{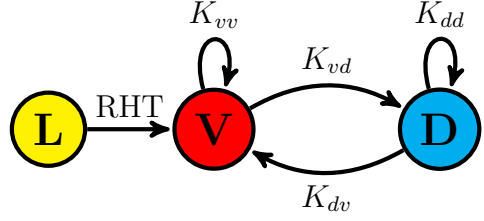}
\caption{ Subpopulations and coupling in the SCN. Light input comes into the sensor cells in the ventral SCN through the rentino-hypothalamic tract (RHT). The majority of the dorsal cells do not receive direct input from the RHT but are bidirectionally coupled to the ventral sensor cells. Coupling terms are labeled as in Eqs.~\ref{Eq:Full}. \label{FIG:EXP}}
\end{figure}

In mammals light input comes in through the eyes and is channeled to the SCN along the retinohypothamic tract (RHT) \cite{Meijer2003}. It is remarkable that the core circadian clock receives light information through a direct pathway from the eyes, which underscores the importance of light in entraining mammalian circadian rhythms. However, only a fraction of the clock cells in the SCN receive light input directly with the majority of cells receiving input in the ventral region \cite{Meijer2003}. Therefore, for the purposes of our model we split the SCN into ventral and dorsal phase clusters and allow light-input into only the ventral population (Fig.~\ref{FIG:EXP}). 

Coupling between clock neurons in the SCN is mediated by a large suite of neurotransmitters \cite{Lee2013}. In this work we focus on the functional coupling between the regions, although it may assist the reader to give an interpretation of the coupling in reference to two predominant neurotransmitters in mammalian circadian rhythms: vasoactive intestinal polypeptide (VIP) and $\gamma$-aminobutyric acid (GABA) whose properties have been characterized experimentally. Perhaps, the best understood coupling agent in the SCN is VIP which is released by the ventral population and is received by all the oscillators \cite{An2012}. VIP is known to be a synchronizing force in the SCN (phase attractive) \cite{Maywood2006}. Recent experimental results \cite{Myung2015, Farajnia2014a} and detailed mathematical modeling \cite{Dewoskin2015} suggest that GABA mediated coupling is more subtle. GABA is released and received by all or nearly all clock neurons in the SCN and has been identified as both a synchronizing \cite{Liu2000} and desynchronizing \cite{Freeman2013, Aton2006} agent among clock cells, although recent evidence suggests these properties vary spatially in the SCN \cite{Myung2015, Dewoskin2015}.  Evidence suggests that both VIP and GABA are involved in the communication of phase shifts between the ventral and dorsal SCN as well as the storage of seasonal day-length information in the SCN \cite{Mintz2002, Myung2015, Dewoskin2015, Cao2013}

Here we assume the combined action of VIP and GABA act to modulate the strength of the coupling between the ventral and dorsal phase clusters in the SCN. 

This conceptional model, summarized in Fig.~\ref{FIG:EXP},  may be translated into a coupled phase oscillator system,
\begin{subequations}
\label{Eq:Full}
\begin{align}
\frac{d \phi_k^v}{dt}&=\omega_k^v+\frac{K_{vv}}{M_v}\sum_{j=1}^{M_v} \sin(\phi_j^v-\phi_k^v) +\frac{K_{dv}}{M_d}\sum_{j=1}^{M_d} \sin(\phi_j^d-\phi_k^v) +  \epsilon Q(\phi_k^v) \delta(t') +\sqrt{D} \eta^v_k(t) \\
\frac{d\phi^d_k}{dt}&=\omega_k^d+\frac{K_{dd}}{M_d}\sum_{j=1}^{M_d} \sin(\phi_j^d-\phi_k^v)+\frac{K_{vd}}{M_v} \sum_{j=1}^{M_v} \sin(\phi_j^v-\phi_k^d)  +\sqrt{D} \eta^d_k(t) , 
\end{align}
\end{subequations}
where $Q(\phi)$ gives the microscopic phase response curve of the ventral oscillators to light and $\eta^{v,d}_k$ defines a white noise process i.e. ($\langle \eta_k(t)\rangle=0$ and $\langle \eta_k(t) \eta_l(t') \rangle=2\delta_{kl} \delta(t-t')$). The $\epsilon$ factor scales the microscopic phase response curve and is taken to be a small parameter.  The coupling strengths are given as $K_{from,to}$ and we let $M_{v,d}$ indicate the total number of oscillators which fall into the ventral and dorsal phase clusters. Finally, we define $q=M_v/(M_v+M_d)$ to be the fraction of ventral (sensing) oscillators in the population and $p=1-q$. 

We allow the oscillators within the ventral and dorsal regions to be heterogeneous in their intrinsic frequencies and assume each cluster has a Cauchy (Lorentzian) distribution of frequencies, 
\begin{equation}
g_{v,d}(\omega)=\frac{1}{\pi} \frac{\gamma}{(\omega-\omega_{0}^{v,d})^2+\hat{\gamma}^2},
\label{Eq:Lorentz}
\end{equation}
with the mean frequency $\omega_0^{v,d}$ and dispersion parameter $\hat{\gamma}$. 

In addition to daily timekeeping, the SCN is also responsible for storing seasonal day-length information \cite{Coomans2015, Meijer2007, Sumova1995}. It has been shown by several experimental groups that the phase difference between the dorsal and ventral clusters grows with the seasonal day-length, making this a leading hypothesis for how seasonal information is encoded in the SCN\cite{Myung2015, Evans2013}. Additionally, it has been suggested that the physiological root of this seasonal variation in the phase difference is alterations in the coupling forces in the SCN \cite{Myung2015, Dewoskin2015, Buijink2016}. Thus, we incorporate seasonal effects (day-lengths) into our model by allowing the coupling strengths $K_{vd}$ and $K_{dv}$ to vary with the seasonal day-length, as these coupling terms will be seen to control the phase difference between the ventral and dorsal populations. 

\subsubsection*{Macroscopic Model}

The model for the mammalian SCN as given in Eq.~\ref{Eq:Full} gives a high-dimensional representation of the dynamical state of the circadian rhythm as $M_v+M_d= \mathcal{O}{(10^4)}$. This high dimensional representation of the system makes analytical analysis of the light-response difficult. Therefore, we make use of the recently developed $m^2$ ansatz to derive a low-dimensional macroscopic model for the ventral and dorsal phase clusters \cite{Hannay2018}. Crucially, the use of this ansatz may be justified through comparison of the core assumption with experimental data on the phase distribution of cellular oscillators in the mammalian SCN \cite{Hannay2018}. 

First, we define the Daido order parameters  \cite{Daido1996,Daido1993} for the ventral and dorsal phase clusters as:
\begin{subequations}
\begin{align}
&Z_n^v=\frac{1}{M_v} \sum_{j=1}^{M_v} e^{in \phi_j^v}, \\
&Z_n^d=\frac{1}{M_d} \sum_{j=1}^{M_d} e^{in \phi_j^d},
\end{align}
\end{subequations}
where $n \in \mathbb{Z}$. The special case of $n=1$ gives the classical Kuramoto order parameter $Z_1^{v,d}=R_1^{v,d}e^{i \psi^{v,d}_1}$, where $R$ is known as the phase coherence and gives a measure of the overall synchrony in the population: when $R_1=1$ the population is phase locked in perfect synchrony and $R_1=0$ when the population is completely desynchronized. Additionally, $\psi_1$ gives the mean phase of the population. For simplicity of notation we will drop the subscript for the Kuramoto order parameters, i.e. $Z_1^{v,d}=Z^{v,d}=R^{v,d}e^{i \psi^{v,d}}$. Using these order parameter definitions we may rewrite Eqs.~\ref{Eq:Full} as,
\begin{subequations}
\label{Eq:Full2}
\begin{align}
&\frac{d \phi_k^v}{dt}=\omega_k^v+K_{vv}\Im[e^{-i\phi_k^v} Z^v]+K_{dv}\Im[e^{-i\phi_k^v} Z^d]  +\sqrt{D} \eta^v_k(t)\\
&\frac{d\phi^d_k}{dt}=\omega_k^d+K_{dd}\Im[e^{-i \phi^d_k} Z^d]+K_{vd}\Im[e^{-i \phi^d_k}Z^v] +\sqrt{D} \eta^d_k(t),
\end{align}
\end{subequations}
with $\Im$ denoting the imaginary part of the expression. In the continuum limit $M_{s,n} \rightarrow \infty$ Eqs.~\ref{Eq:Full2} give rise to continuity equations for the phase density functions $f^{v,d}(\omega, \phi,t)$,
\begin{subequations}
\label{Eq:Cont}
\begin{align}
&\frac{\partial f^{v,d}}{\partial t}+ \frac{\partial}{\partial \phi}(f^{v,d} W_{v,d})+D \frac{\partial^2 f^{v,d}}{\partial \phi^2}=0 \\
& W_{v}=\omega+K_{vv}\Im[e^{-i \phi}Z^v]+K_{dv}\Im[e^{-i\phi} Z^d] \\
&W_d=\omega+K_{dd}\Im[e^{-i\phi} Z^d]+K_{vd}\Im[e^{-i \phi} Z^v]. 
\end{align} 
\end{subequations}
Let us consider the Fourier series representation of the phase density functions $f^{v,d}(\omega, \phi,t)$,
\begin{align}
\label{Eq:FS}
&f^{v,d}= \frac{g_{v,d}(\omega)}{2 \pi} \left(1+\left[\sum_{k=1}^{\infty}A^{v,d}_k(\omega,t) e^{ik \phi}+ \text{c.c}  \right]    \right),
\end{align}
where c.c stands for the complex conjugate. Inserting the Fourier series representation into the continuity equation gives a system for the Fourier coefficients $A^{v,d}_k(\omega,t)$,
\begin{subequations}
\label{Eq:MC}
\begin{align}
&\frac{(A^v_k)'}{k}+(i\omega+Dk) A^v_k+\frac{K_{vv}}{2}\left[\tilde{Z}^v A_{k+1}^v-Z^v A_{k-1}^v \right]  + \frac{K_{dv}}{2}\left[ \tilde{Z}^d A_{k+1}^v-Z^d A_{k-1}^v   \right]=0, \\
&\frac{(A^d_k)'}{k}+(i\omega+Dk) A^d_k+\frac{K_{dd}}{2}\left[\tilde{Z}^d A_{k+1}^d-Z^d A_{k-1}^d \right]+ \frac{K_{vd}}{2}\left[ \tilde{Z}^v A_{k+1}^d-Z^v A_{k-1}^d   \right]=0,
\end{align}
\end{subequations}
with the tilde representing the complex conjugate. In the continuum limit the Daido order parameters $Z_{n}^{v,d}$ are given by,

\begin{align}
Z^{v,d}_n(t)&=\int_0^{2\pi} \int_{-\infty}^{\infty} f^{v,d}(\omega, \phi,t) e^{i n\phi} d\omega d \phi =\int_{-\infty}^{\infty} \tilde{A}_n^{v,d}(\omega,t) g_{v,d}(\omega) d\omega, \label{Eq:ContOP2}
\end{align}
using that all the terms in the Fourier series integrate to zero except the $n=k$ term. Further, since we approximate the natural frequency distribution as a Cauchy/Lorentzian distribution (Eq.~\ref{Eq:Lorentz}) we may evaluate the integral (Eq.~\ref{Eq:ContOP2}) under the assumption that $A_k(\omega,t)$ may be analytically continued into the complex $\omega$ plane \cite{Ott2008}. Thus we have that,
\begin{align}
Z^{v,d}_n(t)&=\tilde{A}_n^{v,d}(\omega_0^{v,d}-i\hat{\gamma},t).
\end{align}
This substitution into Eqs.~\ref{Eq:MC} gives,
\begin{subequations}
\begin{align}
&\frac{(Z_k^v)'}{k}=i\omega Z^v_k-\hat{\gamma} Z^v_k+\frac{K_{vv}}{2}\left[Z^v Z^v_{k-1}-\tilde{Z}^v Z^v_{k+1}  \right] +\frac{K_{dv}}{2}\left[ Z^d Z^v_{k-1}-\tilde{Z}^d) Z^v_{k+1} \right] -Dk Z^v_k\\
&\frac{(Z_k^d)'}{k}=i\omega Z^d_k-\hat{\gamma} Z^d_k+\frac{K_{dd}}{2}\left[Z^d Z^d_{k-1}-\tilde{Z}^d) Z^d_{k+1}  \right] +\frac{K_{vd}}{2}\left[ Z^v Z^d_{k-1}-\tilde{Z}^v Z^d_{k+1} \right] -Dk Z^d_k
\end{align}
\end{subequations}
Finally, we consider the system with $k=1$ and apply the $m^2$ ansatz  ($ Z_m=|Z_1|^{m^2-m} Z_1^{m}$ or $R_m=R_1^{m^2}$, $\psi_m=m\psi_1$). Applying the $m^2$ ansatz and separation into the real and imaginary parts of the expressions gives a four dimension system describing the phase coherence $R_{v,d}$ and mean phase $\psi_{v,d}$ of each cluster. However, with a change of variables $\theta=\psi_d-\psi_v$ (``phase gap") and letting $\Delta \omega=\bar{\omega}_d-\bar{\omega}_v$ and $\gamma=\hat{\gamma}+D$ we arrive at a three dimensional system of equations:
\begin{subequations}
\label{Eq:HAReducedGen}
\begin{align} 
\dot{R_v}&=-\gamma_v R_v+ \frac{K_{vv}}{2} R_v(1-R_v^4) +\frac{K_{dv}}{2} R_d(1-R_v^4) \cos(\theta) \label{Eq:RvEquation}\\
\dot{R}_d&=-\gamma_d R_d \frac{K_{dd}}{2}R_d(1-R_d^4) + \frac{K_{vd}}{2}R_v(1-R_d^4)\cos(\theta) \\
\dot{\theta}&=\Delta \omega -G \sin(\theta)  \\
G&=\frac{R_v R_d}{2}\left[ K_{vd}\left (R_d^2+\frac{1}{R_d^2} \right)+K_{dv}\left(R_v^2+\frac{1}{R_v^2}\right)  \right ]
 \label{Eq:Theta}
\end{align}
\end{subequations}
By setting $\bar{\omega}=q\omega_0^v+p\omega^d_0$, we can define $\Omega=q\dot{\psi}_v+p\dot{\psi}_d$ as the collective frequency of the system in a synchronous state,
\begin{subequations}
\label{Eq:Omega}
\begin{align}
&\Omega=\bar{\omega}+H \sin(\theta)\\
&H=\frac{R_v R_d}{2} \left[qK_{dv}\left(R_v^2+\frac{1}{R_v^2}\right)-pK_{vd}\left(R_d^2+\frac{1}{R_d^2}\right)  \right] \label{Eq:Heq}
\end{align}
\end{subequations}
In order to facilitate our analysis we define a default parameter set for this model given in Table~\ref{TB:Param}. Under this parameter set Eqs.~\ref{Eq:HAReducedGen} evolve to a fixed point $(R_v^*, R_d^*, \theta^*)$ with a collective frequency $\Omega^*$-we will use starred quantities refer to fixed-point solutions. 

Finally, we note that experimental evidence has shown that the phase gap between the dorsal and ventral populations is typically a small variable $\theta^*\in [0, 0.5]$ radians for photoperiods in the range of 6-18 hours of light \cite{Myung2015}, although it may grow considerably when mice are kept in twenty hours or more of light each day \cite{Evans2013}. 

\begin{table}
\centering
\begin{tabular}{|c|c|}
\hline
	\textbf{Parameter} & \textbf{Value} \\
	\hline 
	$\bar{\omega}_v$ & $2 \pi/24.5$ \\
\hline
	$\bar{\omega}_d$ & $2 \pi/23.5$\\
\hline
	$\gamma$ & 0.024\\
\hline
	$K_{vv}$ & 0.095\\
\hline
	$K_{dd}$ & 0.07\\
\hline
$K_{vd}$ & $\alpha K_{dv}$\\
\hline
$K_{dv}$ & 0.05 \\
\hline
$\alpha$ & 2.0\\
\hline
$q$ & 0.5\\
\hline
\end{tabular}
\caption{Parameter sets used for numerical simulations in the main text (default parameter set). These parameters give steady state values of $R_v^*=0.81$, $R_d^*=0.84$, $\theta^*=0.06$ \label{TB:Param} } 
\end{table}

\subsection*{Collective Phase Response Curves}

\subsubsection*{Components of the Collective Phase Response Curve}

The collective phase response to a stimulus may be defined by the shift in the mean-phase $\psi=Arg(Z)$ induced by the light perturbation. For a brief ( Dirac $\delta$ function) stimulus we may break the collective phase shift into two components \cite{Levnajic2010}:
\begin{enumerate}
\item The prompt phase shift ($\Delta_0$) induced at $t=t'$ the instant the stimulus is applied. 
\item The relaxation phase shift ($\Delta_R$) which results from any phase shifts induced as the system relaxes to its asymptotic state. 
\end{enumerate}
The collective phase shift ($\Delta_{\infty}$) is then given by,
\begin{align}
\label{Eq:baseDeltaInf}
\Delta_{\infty}=\Delta_0+\Delta_{R}=Arg\left(\frac{\bar{Z}}{Z_0} \right)+\Delta_R
\end{align}
where we define $Z_0$ as the order parameter just prior to the perturbation and $\bar{Z}$ as the order parameter just after the perturbation. Notationally, barred quantities will refer to the quantity just after the perturbation is applied.  

It is also useful to define the amplitude resetting curve $\Lambda$ as a measure of the perturbations transient effect on the amplitude of the collective rhythm,
\begin{align}
\Lambda=\left|\frac{\bar{R}}{R_0}  \right|. \label{Eq:DefAmpRes}
\end{align}
Given the assumed stability of the limit cycle, perturbations of the amplitude $R$ are expected to decay, thus the amplitude resetting is defined in terms of the initial amplitude reduction imposed on the system.  

\subsubsection*{Single Population Case}
We first consider the collective phase response for a single population of oscillators, that is,  we consider the case where all oscillators in the population receive the light stimulus. These results will aid our consideration of the two population case, as they can be used to describe the phase shift in the ventral oscillator population. For a general phase response curve $Q(\phi)$ we have previously derived an asymptotic formula for the collective phase response of a single population \cite{Hannay2015} of Kuramoto-Sakaguchi oscillators making use of the Ott-Antonsen formalism. In this section we adapt those results to study phase shifts in a population which follows the $m^2$ ansatz as has been found in experimental measurements of the SCN phase distribution \cite{Hannay2018}. For times close to the perturbation $t\approx t'$ we may approximate the single population continuity equation as,
\begin{align}
\frac{\partial f}{\partial t} +\frac{\partial}{\partial \phi}\left[ \epsilon f(\omega, \phi, t) Q(\phi(t)) \delta(t) \right].
\label{Eq:ContSQ}
\end{align}
We have previously shown that for small $\epsilon$ we may approximate the solution of Eq.~\ref{Eq:ContSQ} as,
\begin{align}
\bar{f}(\omega, \phi, t)=f(\omega, \phi-\epsilon Q(\phi))e^{-\epsilon Q(\phi-\epsilon Q(\phi))}, \label{Eq:MCS}
\end{align}
by employing the method of characteristics. Expanding Eq.~\ref{Eq:MCS} to leading order in $\epsilon$, multiplying by $e^{i\phi}$ and integrating with respect to $\phi$ and $\omega$ gives an expression relating the order parameter after the perturbation $\bar{Z}$ to the order parameter just prior to the perturbation $Z_0$. 
\begin{align}
\bar{Z} \approx Z_0 +i \epsilon \int_{-\pi}^{\pi} \int_{-\infty}^{\infty} f(\omega, \phi,t) Q(\phi) e^{i \phi} d\omega d\phi. \label{Eq:ZbarZ01}
\end{align}
 Now, we replace the microscopic PRC $Q(\phi)$ with its Fourier Series representation,
\begin{subequations}
\label{Eq:FourierSeries}
\begin{align}
Q(\phi)&=\frac{A_0}{2} +\sum_{n=1}^{\infty} A_n e^{i n \phi}+\tilde{A}_n e^{-i n \phi}= \frac{A_0}{2} + \sum_{n=1}^{\infty} a_n \sin(n\phi)+b_n \cos(n\phi).
\end{align}
\end{subequations}
Substitution of the Fourier series representation into Eq.~\ref{Eq:ZbarZ01}) and applying the definition of the Daido order parameters gives,
\begin{align}
\bar{Z}=Z_0+i \epsilon \left[ \frac{A_0}{2}Z_0+\sum_{n=1}^{\infty} A_n Z_{n+1}  +\tilde{A}_n Z_{n-1}         \right].
\end{align}
Now we apply the $m^2$ ansatz [$R_m=R^{m^2}$,  $\psi_m=m \psi$] to arrive at an expression for $\bar{Z}$ in terms of $Z_0$, 
\begin{subequations}
\begin{align}
&\bar{Z}=Z_0(1+i\epsilon \hat{Q}(\psi, R) ) \\ 
&\hat{Q}=\frac{A_0}{2}+\frac{1}{R} \sum_{n=1}^{\infty} A_n R^{(n+1)^2} e^{in\psi} + \tilde{A}_n R^{(n-1)^2} e^{-in\psi}.
\end{align}
\label{Eq:ZbarS}
\end{subequations}
By applying Eq.~\ref{Eq:baseDeltaInf} and Eq.~\ref{Eq:DefAmpRes} we may derive expressions for the prompt resetting $\Delta_0$ and the amplitude response curve $\Lambda$ respectively,  
\begin{subequations}
\label{Eq:VentralOnly}
\begin{align}
&\Delta_0=Arg\left(\frac{\bar{Z}}{Z_0}  \right)=\epsilon Re[\hat{Q}(\psi)] \label{Eq:Delta0S}\\
&\Lambda=\left|\frac{\bar{R}}{R_0} \right|=1-\epsilon Im[\hat{Q}(\psi)].  \label{Eq:LambdaS}
\end{align} 
\end{subequations}
The real part of $\hat{Q}(\psi)$ can be compactly expressed in terms of the Fourier series for the microscopic PRC Eqs.~\ref{Eq:FourierSeries},
\begin{subequations}
\label{Eq:VentralReal}
\begin{align}
&Re[\hat{Q}]=\frac{A_0}{2}+\sum_{k=1}^{\infty}f_k(R)[a_k \sin(k\psi)+b_k \cos(k\psi)] \label{Eq:realQhat}\\
&f_k(R)=\frac{1}{2} R^{k^2}(R^{2k} +\frac{1}{R^{2k}}). 
\end{align}
\end{subequations}
From these expressions we can see the principal effect of the phase distribution on the shape of the PRC is to re-weight the Fourier harmonics according to $f_k(R)$. As $R \rightarrow 1$ we have that $f_k \rightarrow 1$ and the collective and microscopic prompt phase response curves coincide. However, when $R<1$ the first harmonic of the microscopic phase response curve is amplified like $R^3+\frac{1}{R}$ while the higher harmonics are damped out by higher powers of $R$. 

The imaginary part of $\hat{Q}(\psi)$ can be expressed as,
\begin{subequations}
\label{Eq:VentralImag}
\begin{align}
&Im[\hat{Q}]= \sum_{k=1}^{\infty}g_k(R)[a_k \cos(k\psi)-b_k \sin(k\psi)] \\ 
&g_k(R)=\frac{1}{2} R^{k^2}(\frac{1}{R^{2k}}-R^{2k}). \label{Eq:imagQhat}
\end{align}
\end{subequations}
In this case we see that the modulation term $g_k(R)$ goes to zero as $R \rightarrow 1$ meaning the amplitude is unaffected by the stimulus in this limit. Additionally we observe that,
\begin{equation}
a_k \cos(k\psi)-b_k \sin(k\psi)  \propto \frac{dQ}{d\psi}, \nonumber
\end{equation}
so we expect the amplitude shifts $\Lambda$ to be greatest around the zeros of the microscopic phase response curve, with transient increases in $R$ around stable points and decreases around unstable zeros.

\subsubsection*{Application to Light PRCs}
The mammalian phase response curve to light has been characterized over a large variety of species and conditions \cite{StHilaire2012, Johnson1999}. Generally, the circadian rhythm shows small sensitivity to light during the subjective day, phase delays during the early subjective night and phase advances in the late subjective night \cite{Pittendrigh1988a}.  Although we expect our results to hold more generally we will focus our attention on phase response curves with this general shape. 

The single population results derived in the last section indicate that for microscopic PRC's dominated by their first harmonic, as are commonly assumed in the circadian literature \cite{Pittendrigh1991, Abraham2010}, a general amplification in the phase response is expected as the oscillators are more dispersed in phase or have a reduced amplitude (either through weaker coupling or greater frequency heterogeneity in the population). This expectation has surfaced in the circadian literature under a variety of guises in the context of both phenomenological and biochemically motivated models \cite{Abraham2010}. 

\begin{figure}
 \includegraphics[scale=1]{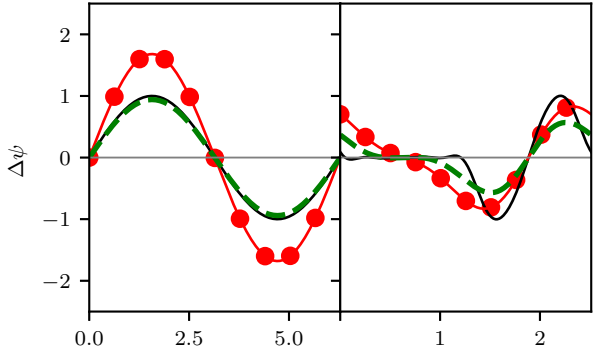}
 \caption{ Collective phase response curves for two assumed microscopic phase response curves: (left) Simple phase response curve $Q(\psi)=\sin(\psi)$ , (right) Light-like PRC shape $Q(\psi)=H(-\sin(\psi))\sin(-2\psi)$. Shown for three phase coherence $R$ values: R=1 (solid black) R=0.6 (dashed green) and R=0.3 (red circles).   \label{Fig:Thumb} }
 \label{Fig:Delta0}
 \end{figure}

These observations have led to the formation of a rule of thumb in the circadian literature that the phase response of a limit cycle oscillator should increase when the collective amplitude/phase coherence decreases \cite{Abraham2010}. However,  our analysis shows this prediction will only hold when the underlying individual/microscopic phase response curve is dominated by its first harmonic-otherwise a overall reduction in the amplitude may be observed \cite{Hannay2015} (Eq.~\ref{Eq:VentralReal}). An example of this effect is shown in Fig.~\ref{Fig:Thumb} for a first harmonic phase response curve and a light-like PRC shape. The first harmonic curve shows a uniform increase in amplitude as the phase coherence of the sensing population decreases, whereas the light-like PRC shows a initial decrease in amplitude as the higher harmonics are dissipated. 

The general shape of the mammalian phase response curve to light has significant power at higher harmonics \cite{StHilaire2012}. Therefore,  the rule of thumb does not necessarily apply in this case. For example, the phase coherence of the ventral (sensor) population is known to decrease with increasing day-length  \cite{Brown2009} which leads to the expectation of increasing amplitude in the response to light. However, as previously noted  the opposite trend has been observed experimentally where organisms entrained to longer day-lengths show \emph{decreased} sensitivity to light-pulses \cite{VDL2009}. 

This reduction in amplitude of the collective phase response, in spite of a reduction in the phase coherence of the sensing population, may be at least partially explained when the higher harmonics in the microscopic phase response to light are taken into consideration. This effect has been noted in the course of simulations \cite{VDL2009} and is readily explained by the theory given here using the $m^2$ ansatz and detailed previously for cases adhering to the Ott-Antonsen ansatz \cite{Hannay2015}. 

In the following section we investigate the effects of only having a fraction of the total population shift in response to a light pulse. This analysis reveals an additional effect which allows the amplitude of the collective phase response to be modulated depending on  the degree of asymmetry in the coupling between the ventral and dorsal populations. 
 
\subsubsection*{Two Population Case}
\label{SEC:Collective}
We now consider the collective phase response of the circadian model Eqs.~\ref{Eq:HAReducedGen} to brief light pulses using that $\Delta_{\infty}=\Delta_0+\Delta_R$. In the first section we compute the prompt phase shifting behavior $\Delta_0$ for the circadian model with a subset of sensor cells and observe the effects on the initial phase shifting behavior ($\Delta_0$). In the next subsection, we present a perturbation technique to determine the relaxation phase shift ($\Delta_R$).  

In the figures for this section we consider a microscopic phase response curve $Q(\psi)$ which is fit to experimental measurements of the human phase response curve to brief light pulses \cite{StHilaire2012}. 

\subsection*{Prompt Resetting $\Delta_0$}
We begin by studying the prompt phase shifting curve $\Delta_0$ for the circadian model. By applying Eq.~\ref{Eq:ZbarS} and using that the dorsal population is unaffected by the perturbation, we find the order parameter just after the perturbation $\bar{Z}=q \bar{Z}_v+p \bar{Z}_d=q Z_v(1+i\epsilon \hat{Q}(\psi_v))+p Z_d$. Therefore the prompt phase shift $\Delta_0$ for the SCN model can be derived as:
\begin{subequations}
\begin{align}
\Delta_0&=Arg\left(\frac{\bar{Z}}{Z}\right)=Arg\left(\frac{q Z_v(1+i\epsilon\hat{Q}(\psi_v))+ p Z_d}{q Z_v+p Z_d}  \right) \\
&=Arg\left(1+ i \epsilon \mu \hat{Q}(\psi_v)    \right) , \text{    } \mu=\frac{R_v}{R_v+\eta R_d e^{i \theta}}  \\
&=\arctan\left(\frac{\epsilon Re[\mu \hat{Q}(\psi_v)]}{1-\epsilon Im[\hat{Q}(\psi_v)]} \right) \\
&=\epsilon Re[\mu \hat{Q}(\psi_v)] +\mathcal{O}(\epsilon^2) \label{Eq:Delta0Unexpanded}
\end{align}
\end{subequations}
where $\eta=p/q$ is the ratio of dorsal to ventral (sensors) in the population. We may now expand Eq.~\ref{Eq:Delta0Unexpanded}, 
\begin{subequations}
\label{Eq:Delta0}
\begin{align}
C&=\frac{R_v[R_v+ R_d \eta \cos(\theta)]}{R_v^2+2 R_v R_d \eta \cos(\theta)+ R_d^2 \eta^2}\label{Eq:CDef} \\
D&=\frac{R_v R_d \eta \sin(\theta)}{R_v^2+2 R_v R_d \eta \cos(\theta)+ R_d^2 \eta^2} \\
\Delta_0&=C \Delta_0^v+D(\Lambda_v-1).
\end{align}
\end{subequations}
Which gives an analytical expression for the prompt resetting in our system using our expressions for $\Delta_0^v$ and $\Lambda_v$ for a single population of oscillators (Eqs.~\ref{Eq:VentralOnly}). We note that Eq.~\ref{Eq:Delta0} has the expected limits: As $\eta \rightarrow 0$, $\Delta_0 \rightarrow \Delta_0^v$ and the system converges to the behavior of a single population of oscillators, in addition as the dorsal population grows $\eta \rightarrow \infty$ we see that $\Delta_0 \rightarrow 0$ causing the system to become unresponsive to perturbations. This analytical approximation gives an accurate approximation for the prompt resetting curve when compared with numerical simulations (Fig.~\ref{Fig:Delta0}).

It is interesting to note the differences between our system and a single population of oscillators which all shift in response to the stimulus. In the two population system the damping of higher harmonics in the microscopic PRC is also observed, however we additionally see a decrease in the initial shift as a function of the fraction of oscillators which receive the light pulse. We note that under the assumption that $\theta \approx 0$ we may approximate Eq.~\ref{Eq:CDef} as,
\begin{equation}
C \approx \frac{q R_v}{q R_v+p R_d} \approx q, \label{Eq:SimpleC}
\end{equation} 
giving the intuitive result that the overall amplitude of the initial response scales with the fraction of ventral sensor cells in the population $q$. 

In addition, unlike the single population case we see the prompt phase response curve depends on the amplitude response function $\Lambda_v$. This dependence leads to a slight change in the zeros (entrainment points) of the prompt PRC when compared to the microscopic PRC since $\Lambda_v$ will be largest about the zeros of $Q$. Moreover, this effect is dependent on having a non-zero phase gap between the two populations ($\theta \neq 0$).  
\begin{figure}
 \includegraphics[scale=1]{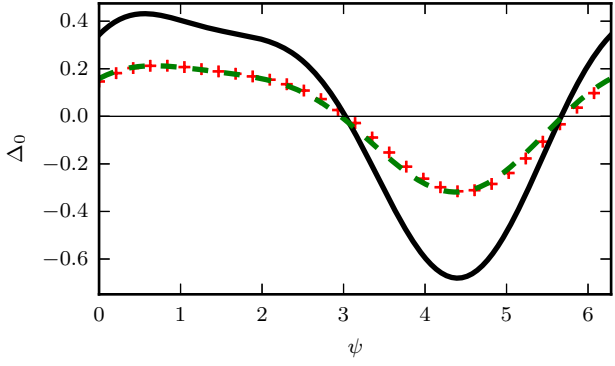}
 \caption{ Prompt Resetting Curve with the fraction of sensors in the population $q=0.50$ and the default parameter values. The microscopic phase response curve (solid black) is fit to the human PRC to a brief light pulse, direct numerical simulations of Eq.~\ref{Eq:Full} with $N=10^4$ (red crosses) and the theoretical prediction Eq.~\ref{Eq:Delta0} (dashed green). }
 \label{Fig:Delta0}
 \end{figure}

\subsubsection*{Relaxation Phase Shift $\Delta_R$}
We now consider the relaxation shift $\Delta_R$ which describes the phase shift induced during the return of the system to equilibrium following a perturbation . For the single population case this relaxation phase shift was directly computable from the Ott-Antonsen equations describing the collective dynamics \cite{Levnajic2010}. This computation relied on the relatively simple spiral isochrons of the Kuramoto-Sakagucki coupling scheme \cite{Hannay2015, ArkadyPikovskyMichaelRosenblum2003, Levnajic2010}.

For the circadian model presented here this quantity can no longer be easily computed by direct integration. However, the relaxation phase shift occurs during the transient decay of the system back to the dynamical fixed point $(R_v^*, R_d^*, \theta^*)$ of the macroscopic model. The collective frequency of the unperturbed system is given by $\Omega^*=\Omega(R_v^*, R_d^*, \theta^*)$. Therefore, the phase shift induced as the system relaxes back to the equilibrium state is given as,
\begin{subequations}
\label{Eq:DeltaRStart}
\begin{align}
\Delta_R&=\int_0^{\infty} \Omega(R_v(t), R_d(t), \theta(t))-\Omega^* dt \\
&=\int_0^{\infty} \Delta \Omega(t) dt
\end{align}
\end{subequations}
Thus, we may calculate the relaxation phase shift by integrating the frequency mismatch between the perturbed system and the steady state system along the trajectory of the system as it returns to equilibrium. The relaxation trajectory may be approximated by a perturbation about the fixed point under the assumption the light-pulse does not induce a large deviation from $(R_v^*, R_n^*, \theta^*)$. We set,
\begin{subequations}
\label{Eq:PertSeries}
\begin{align}
&R_v(t)=R_v^*+\sigma R_v^1(t)+\mathcal{O}(\sigma^2) \\
&R_d(t)=R_d^*+\sigma R_d^1(t)+\mathcal{O}(\sigma^2) \\
&\theta(t)=\theta^*+\sigma \theta^1(t)+\mathcal{O}(\sigma^2),
\end{align}
\end{subequations}
where $\sigma$ is a small parameter and with initial conditions $R_v(0)=R_v^*+\Delta R_v$, $R_d(0)=R_d^*$, $\theta(0)=\theta^*+\Delta \theta$ using that the dorsal population is initially unaffected by the light stimulus. The initial changes in $R_v$ and $\theta$ can be written in terms of the prompt phase and amplitude response curves for the ventral population: $\Delta R_v=R_v (1-\Lambda_v)$ and $\Delta \theta=(\psi_d-\bar{\psi}_v)-(\psi_d-\psi_v)=\psi_v-\bar{\psi}_v=-\Delta_0^v$. 

\begin{figure}
\centering
\includegraphics[scale=1.00]{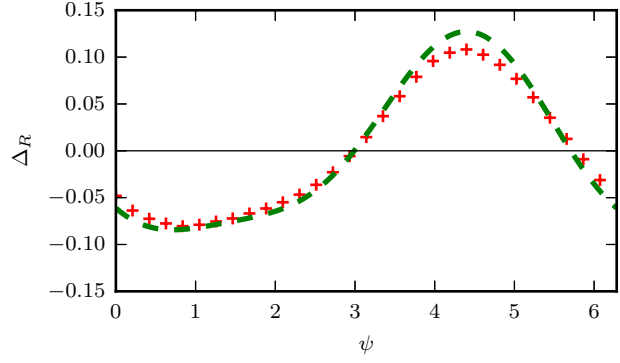}
\caption{ Relaxation phase response curve $\Delta_R$ using a first order perturbation series to calculate $(A,B)$ in Eq.~\ref{Eq:RelaxGen} (dotted green) versus numerical simulation(red crosses) for the default parameter values.  \label{Fig:Relax}} 
\end{figure}

Therefore, the leading order terms in $\sigma$ for the relaxation phase shift is given by,
\begin{align}
\Delta_R\approx A (1-\Lambda_v)-B \Delta_0^v, \label{Eq:RelaxGen}
\end{align} 
with the (A,B) constants determined by the model parameters. In practice we find the leading order term in $\sigma$ is sufficient to provide a good approximation to the numerical solutions (see Fig.~\ref{Fig:Relax}) although higher order terms may be taken in the perturbation series (Eq.~\ref{Eq:PertSeries}) if additional accuracy is required. We note that $(A,B)$ are a measure of the sensitivity of the collective frequency of the system to perturbations in the amplitude ($R_v$) and phase gap $\theta$ respectively, and are weighted by the stiffness of the system to perturbations in those directions. 

\begin{figure}
\centering
\includegraphics[scale=1.00]{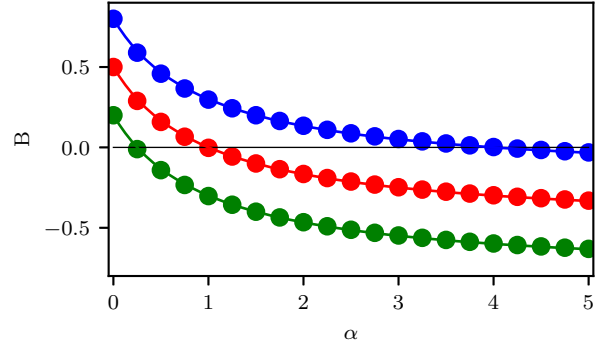}
\caption{ Network resistance to phase shifts $B$ versus  $\alpha=K_{vd}/K_{dv}$ for the first order perturbation theory (circles) with the default parameter values varying $K_{vd}$ and the approximate formula Eq.~\ref{Eq:SimpleDeltaR} (solid line) with $q=0.5$ (red), $q=0.8$ (blue), $q=0.2$ (green).   \label{Fig:BPlot}} 
\end{figure}

To gain intuition of how the circadian model parameters will affect the relaxation phase shifts, we solve for the relaxation terms analytically for a simplified system (see supplementary information). If the amplitude of the ventral and dorsal populations are fixed (without loss of generality let $R_{v,d}=1$) we find that, 
\begin{align}
&B= \frac{q-p\alpha}{1+\alpha}, &\alpha=\frac{K_{vd}}{K_{dv}}. \label{Eq:SimpleDeltaR}
\end{align}
From this simplification we can see that $B \in [-p,q]$ when both coupling terms are positive. Moreover the change occurs at $\alpha=\frac{q}{p}$ from a positive to negative value. Note, for symmetric coupling ( $\alpha=1$ ) between the regions  and $q=p=\frac{1}{2}$ we have $B=0$. In Fig.~\ref{Fig:BPlot} we show the variation of $B$ with $\alpha$ using both the perturbation approach and the simplified formula Eq.~\ref{Eq:SimpleDeltaR}. 

\subsubsection*{Collective Phase Response Curve }
\label{SEC:CollectivePRC}
When the prompt phase shift $\Delta_0$ is combined with the relaxation shift $\Delta_R$ we find the collective phase response curve $\Delta_{\infty}$,
\begin{align}
\label{Eq:cPRC}
\Delta_{\infty}=\Delta_0+\Delta_R=(C-B)\Delta_0^v+(D-A)(1-\Lambda_v),
\end{align}
with the constants $(A,B,C,D)$ as defined in the previous sections. In general, we find this approximate formula provides a good approximation to the numerically determined collective phase response (Fig.~\ref{Fig:cPRC}). To improve our understanding of the role of light-input to the mammalian circadian system we analyze these results further. Of particular importance is the amplitude of the collective phase response as this determines the entrainment range for weakly forced systems \cite{Winfree1980}. 

We first note that the collective PRC for the circadian system carries over many of the trends of the single population model. Namely, we expect that higher harmonics in the microscopic phase response curve will be damped with the first harmonic amplified like $R_v^3+\frac{1}{R_v}$. This transformation in the shape leads to an overall smoothing effect and is tied to the disorder in the underlying population. 

Additionally, we note that the term proportional to the amplitude response curve $(D-A)(1-\Lambda_v)$ is expected to be of comparatively small magnitude as it is proportional to $\frac{1}{R_v}-R_v^3 \rightarrow 0$ as $R_v \rightarrow 1$. Moreover the amplitude response curve reaches its maximum values around the zeros of the individual phase response curve. Thus, the amplitude of the collective phase response curve is largely determined by the $(C-B)\Delta_0^v$ term, while a shift in the entrainment points is determined by the $(D-A)(1-\Lambda_v)$ term. 

From Eq.~\ref{Eq:cPRC} we see the amplitude of the collective PRC is influenced by the sign of the $B$ constant as determined by the relaxation dynamics. Positive values of $B$ indicate the relaxation shift acts to decrease the initial phase shift thus providing a resistance to the phase shift. Negative values of $B$ indicate the initial shift is reinforced/increased during the transient relaxation, while a value near zero indicates the relaxation shift has a small effect on the collective phase response.

The simplified expression in Eq.~\ref{Eq:SimpleDeltaR} allows for intuition on the scale and sign of $B$. In particular we can see the value of $B$ scales with the ratio of the feedforward coupling strength $K_{vd}$ to the feedback strength $K_{dv}$. For a balanced system $K_{vd}=K_{dv}, q=0.5$ we observe that $B=0$, although by varying this ratio of coupling strengths the system can toggle between a resistant/reinforcing behavior to the initial phase shifts.  

For a pure feedforward network, where $K_{dv}=0$, the $B$ term is positive and the effect of having a fractional sensor population on the amplitude of the collective phase response disappears. In this limit the phase shift in the ventral population is imposed on the dorsal population over time. Thus, when $K_{dv}\neq 0$ the non-sensing dorsal population can act as a feedback on the phase shifts and integrate the current phase shift against the past history. 

\begin{figure}
\centering
\includegraphics{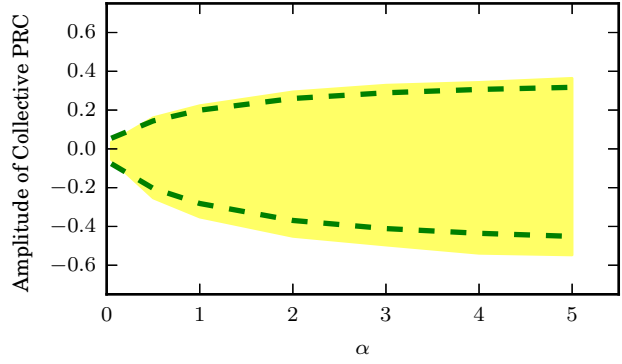}
\caption{ The amplitude of the collective phase response curve as a function of $\alpha=K_{vd}/K_{dv}$. The range of the numerical collective phase response is highlighted in yellow and the theoretical prediction of the amplitude ( Eq.~\ref{Eq:cPRC}) is shown as dotted green lines. \label{Fig:Amp} }
\end{figure}

Therefore, we see that the entrainment range of a two-population system is crucially dependent on the ratio of the coupling strengths between the sensing (ventral) and non-sensing (dorsal) populations (Fig.~\ref{Fig:Amp}). By adjusting the ratio of these coupling strengths the size of the light response may be modulated. 
 
\begin{figure}
\centering
\includegraphics[scale=1.00]{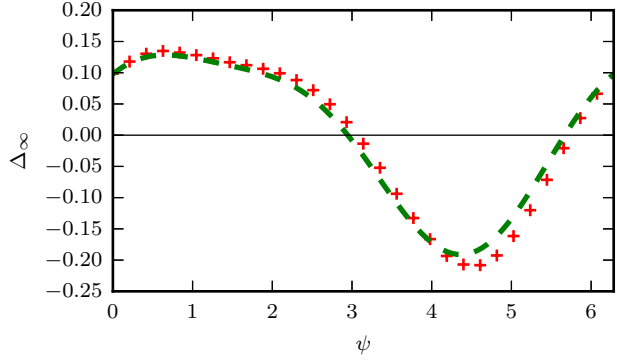}
\caption{ Collective phase response curve \label{Fig:cPRC} for the theoretical curve (Eq.~\ref{Eq:cPRC})  (dotted green) versus numerical simulation (red crosses) for the default parameter values. }
\label{Fig:cPRC}
\end{figure}

\subsection*{Seasonal Effects on Light Resetting}
\label{SEC:Seasonal}
Our study of the collective phase response in a two-population model identified the relative strengths of the intra-population coupling forces as an important factor in determining the amplitude of the phase response. Experimental evidence has found that the amplitude of phase shifts induced by a light-stimulus decreases in mammals entrained to long day-lengths \cite{VDL2009, Ramkisoensing2014}.  This could be explained in our framework by an increase in the network resistance to the phase shift  ($B$), in animals exposed to long day-lengths. 

To evaluate this hypothesis we may check for consistency against two other seasonal light effects on mammalian circadian rhythms: Seasonal encoding and light entrainment after-effects. First, we consider seasonal encoding.  Experimental evidence has indicated that the phase difference between the dorsal and ventral phase clusters grows with the day-length \cite{Myung2015, Evans2013}.  Within our model this corresponds to the variable $|\theta^*|$ growing with the entrained day-length. We make use of the hypothesis that these changes in $\theta$ occur through an adjustment of the couping strengths rather than a change in the intrinsic periods in the ventral/dorsal SCN \cite{Dewoskin2015, Myung2015}. Thus, we consider the intrinsic periods to be constant while allowing the coupling strengths to vary. 

Let $\chi \in [0,1]$ be the photoperiod, or the fraction of the circadian day in which the organism is exposed to light. Considering Eq.~\ref{Eq:HAReducedGen} we see the steady state phase gap $\theta^*$ is given by,
\begin{subequations}
\begin{align}
\theta^*(\chi)&=\arcsin\left( \frac{\Delta \omega}{G(\chi)} \right)\approx \left( \frac{\Delta \omega}{G(\chi)} \right)\\
\frac{d|\theta^*|}{d\chi}&=\frac{-|\Delta \omega| G'(\chi)}{G(\chi)^2} 
\end{align}
\end{subequations}
From this we can see that we must have $G'(\chi)<0$ to allow the absolute value of the phase gap to increase with the photoperiod length. 

Additionally, we observe that $\theta^*$ will only show significant variation for a small range $G(\chi)=\mathcal{O}(\Delta \omega)$ and will asymptote to a small value for $G(\chi)$ outside this range. This nonlinear dependence of the phase gap $\theta^*(\chi)$ with asymptotic values for short photoperiods has been observed in experiments using both \textit{Per2} and \textit{Bmal} circadian phase markers \cite{Myung2015, Evans2013}. 

Moreover, the macroscopic model also identifies a fundamental trade-off as the photoperiod is lengthened. As $G(\chi)$ decreases towards $\Delta \omega$ the phase gap increases quickly. However, a further increase in $\chi$ will cause the system to undergo a bifurcation where the ventral and dorsal regions decouple from one another. Thus, we predict that organisms which show robust seasonal adjustment necessarily must approach a bifurcation to desynchrony/large phase gaps at long photoperiods.  Therefore, organisms which show robust decreases in $G(\chi)$, thereby showing larger changes in the phase gap variable $\theta^*$ as the photoperiod lengthens, will also approach a bifurcation to desynchrony more closely and display rhythm abnormalities. 

In fact, several species of mammals show desynchrony or large phase gaps when perturbed outside their normal photoperiodic range by unnatural lighting conditions (e.g. constant light) \cite{Evans2013, Yan2005, Ohta2005}. Moreover, it has been observed that hamsters which show robust seasonal adjustments to short daylengths have a higher propensity for rhythm abnormalities under constant lighting conditions-illustrating the trade-off identified by our analysis \cite{Evans2012}.

In order to relate these properties to reduced light phase-resetting responses at long photoperiods we now recall a predominant circadian after-effect to light-entrainment in mammals \cite{Myung2015, Pittendrigh1976a}: Mammals entrained under short day-lengths show a transient increase in period when moved to a dark environment. The long day length after-effect works in an opposite direction by inducing a short period with the magnitude of the period change increasing with the entrained day length. In the context of our model this implies that $\frac{d\Omega}{d\chi}>0$ in Eq.~\ref{Eq:Omega}. Expanding this condition gives,
\begin{align}
\frac{d\Omega}{d\chi}=H'(\chi)\sin(\theta^*)+H(\chi)\cos(\theta^*)\frac{d\theta^*}{d\chi}>0.
\end{align}
Applying the assumption that $\theta^*$ is a small variable we may simplify this condition to give,
\begin{align}
\Delta \omega \left[\frac{H'(\chi)G(\chi)-H(\chi)G'(\chi)}{[G(\chi)]^2 } \right] >0, \label{Eq:AE}
\end{align}
and we consider the case that $\Delta \omega >0$ as indicated by experimental evidence \cite{Azzi2017a, Myung2015, Myung2012}. Under the approximation that $R_v\approx R_d$ we have that  $2 H(\chi)\approx qK_{dv}(\chi)-pK_{vd}(\chi)$ and $G(\chi)\approx K_{dv}(\chi)+K_{vd}(\chi)$. This simplification allows us to express our approximation for the network resistance to phase shifts $B$ (Eq.~\ref{Eq:SimpleDeltaR}), in terms of $H$ and $G$,
\begin{align}
&\frac{1}{2} B(\chi)\approx \frac{H(\chi)}{G(\chi)}. 
\end{align}
Thus, taking the derivative with respect to the the photoperiod $\chi$, 
\begin{align}
&\frac{1}{2}\frac{dB}{d\chi}=\frac{ H'(\chi)G(\chi)-H(\chi)G'(\chi)}{[G(\chi)]^2}>0,  \label{Eq:AECond}
\end{align}
we see that the network resistance to phase shifts $B$ will increase with the photoperiod directly from the after-effect condition (Eq.~\ref{Eq:AE}).
This gives the surprising result that seasonal entrainment after-effects and the reduced sensitivity to light-pulses at long photoperiods are intimately related to one another. In fact, the presence of one implies the other in our model. Additionally, we see that the seasonal adjustment condition $G'(\chi)<0$ is consistent with the after-effect and increasing phase shift resistance condition (Eq.~\ref{Eq:AECond}). Therefore, we find the adjustment of the coupling strengths required to explain three predominant light mediated circadian effects are all mutually consistent within our model. 



\section*{Discussion}
\label{SEC:conc}

In this work we focus on phase resetting to light in mammalian rhythms making use of the $m^2$ ansatz to derive a simplified model of the central clock.  The reduced model holds the advantage that the collective variables $(R_v, R_d, \theta)$ all have physiological interpretations and are measurable in experimental treatments. We have focused on the effects of heterogeneity in oscillator frequencies, the shape of the microscopic phase response curve to light, the effects of only a fraction of populations receiving direct light input and variation of the coupling strengths between regions on the phase resetting response. 

Similar to previous work on this subject we find heterogeneity of the population changes the shape, amplitude and zeros of the collective phase response curve \cite{Hannay2015, Levnajic2010}. Moreover, we note these alterations occur through a re-weighting of the Fourier components of the collective phase response with the first harmonic amplified and higher harmonics being damped out as the oscillators spread out in phase. We find examination of only the first harmonic terms may give misleading results when considering phase resetting to light in mammals. 

The effect of having a fraction of the population receive light-input can lead to a reduction in the overall amplitude of the collective phase response. However, we find this effect is dependent on the coupling between the oscillators in the SCN. In a feedforward network, where the sensing ventral cells project more strongly on the non-sensing dorsal cells than the feedback connection, the initial shift induced on the sensing population is largely imposed on the total population over time. Thus, for pure feedforward network architectures the effect of a fractional sensing population is to induce a time delay on the shift of the population mean phase. 

However, when feedback coupling from the dorsal cells to the sensing ventral populations is significant we see the reduction in the initial shift caused by the fractional sensing is retained and even reinforced over time. This leads to the conclusion that the relative coupling strengths between the  ventral and dorsal oscillators allows the system to weight phase shifts induced by light differently. The re-weighting of the coupling strengths between the subpopulations with seasonal changes in day length may act like the aperture on a camera by allowing the sensitivity of the clock to light to vary with the total amount of light input received. In short days the feed-forward connection is weighted more strongly to allow for larger responses when light input is more scarce. In contrast, relatively weighting of the feedback connection more strongly in long days results in a reduction in light sensitivity when more light is received over the course of the day. 

These results may also have applications to the study of aging in the circadian clock. In a similar manner to long-day lengths, older animals tend to show reduced phase coherence and the clock neurons are thought to be more weakly coupled as the animals age \cite{Farajnia2012, Nygard2005, Nygard2006}. Moreover, these aged animals also show reduced phase-shifts in response to light stimuli and slower entrainment to shifted light schedules \cite{Benloucif1997, Sellix2012, Valentinuzzi1997, Biello2009}. Our analysis reveals this reduction in phase shifting capacity may be explained in terms of a reduction in the ratio of the feed-forward to feedback coupling between the ventral and dorsal populations. 

Our results are consistent with current ideas of how seasonal information is encoded the mammalian circadian clock \cite{Myung2015, Evans2016} and provide an explanation for mammals showing reduced phase shifts to light when entrained to long day lengths \cite{VDL2009, Ramkisoensing2014}. Furthermore, our model reveals an intimate connection between seasonal day-length encoding, seasonal entrainment after-effects and the amplitude of the phase-response to light. Additionally, we find the change in coupling strengths with the day-length required to explain each of these phenomena are mutually consistent within our model. 

However, our results remain to be strengthened both from a biological and theoretical standpoint. In order to derive these results we have assumed an all-to-all connectivity between clock cells in the SCN and simple sinusoidal coupling between the oscillators. However, the connectivity in the SCN is known to be much more complex \cite{Abel2016}. An important extension of these results would be to consider phase resetting in a general circadian network building on previous results \cite{Kori2006, Kori2004}. 

From a biological standpoint it remains to be tested whether an increasing resistance to phase shifts under long-day lengths underlies the decreased sensitivity to light pulses for organisms entrained to long day-lengths. Although, this prediction seems testable by measuring initial phase shifts to light and comparing these with the asymptotic phase shifts obtained over long-times. This would be particularly interesting if the effect was seen to vary with the entrained day-length and could provide evidence for a variation in the relative coupling strengths between the ventral and dorsal SCN with seasonal day-length. Recent experimental evidence suggests that the coupling strength, as determined globally in the SCN, decreases with increasing day-lengths which could provide indirect evidence for this hypothesis \cite{Buijink2016}. 



\bibliography{references}

\begin{thebibliography}{10}
\urlstyle{rm}
\expandafter\ifx\csname url\endcsname\relax
  \def\url#1{\texttt{#1}}\fi
\expandafter\ifx\csname urlprefix\endcsname\relax\def\urlprefix{URL }\fi
\expandafter\ifx\csname doiprefix\endcsname\relax\def\doiprefix{DOI: }\fi
\providecommand{\bibinfo}[2]{#2}
\providecommand{\eprint}[2][]{\url{#2}}

\bibitem{Pittendrigh1976a}
\bibinfo{author}{Pittendrigh, C.~S.} \& \bibinfo{author}{Daan, S.}
\newblock \bibinfo{journal}{\bibinfo{title}{{A Functional Analysis of Circadian
  Pacemakers in Nocturnal Rodents I. The Stability and Lability of Spontaneous
  Frequency}}}.
\newblock {\emph{\JournalTitle{Journal of Comparative Physiology-A}}}
  \textbf{\bibinfo{volume}{106}}, \bibinfo{pages}{223--252}
  (\bibinfo{year}{1976}).

\bibitem{Johnson1999}
\bibinfo{author}{Johnson, C.}
\newblock \bibinfo{journal}{\bibinfo{title}{{Forty years of PRCs-What have we
  learned?}}}
\newblock {\emph{\JournalTitle{Chronobiology international}}}
  \textbf{\bibinfo{volume}{16}}, \bibinfo{pages}{711--743}
  (\bibinfo{year}{1999}).

\bibitem{Winfree1980}
\bibinfo{author}{Winfree, A.~T.}
\newblock \emph{\bibinfo{title}{{The Geometry of Biological Time}}}
  (\bibinfo{publisher}{Springer}, \bibinfo{address}{New York},
  \bibinfo{year}{2001}).

\bibitem{ArkadyPikovskyMichaelRosenblum2003}
\bibinfo{author}{Pikovsky, A.}, \bibinfo{author}{Rosenblum, M.} \&
  \bibinfo{author}{Kurths, J.}
\newblock \emph{\bibinfo{title}{{Synchronization : A Universal Concept in
  Nonlinear Sciences}}} (\bibinfo{publisher}{Cambridge University Press},
  \bibinfo{address}{Cambridge, England}, \bibinfo{year}{2004}).

\bibitem{Daan2001}
\bibinfo{author}{Daan, S.} \& \bibinfo{author}{Aschoff, J.}
\newblock \bibinfo{journal}{\bibinfo{title}{{The entrainment of circadian
  systems}}}.
\newblock {\emph{\JournalTitle{Circadian Clocks}}}
  \textbf{\bibinfo{volume}{12}}, \bibinfo{pages}{7–43},
  \doiprefix\url{10.1007/978-1-4615-1201-1{\_}2} (\bibinfo{year}{2001}).

\bibitem{Moore1972}
\bibinfo{author}{Moore, R.~Y.} \& \bibinfo{author}{Eichler, V.~B.}
\newblock \bibinfo{journal}{\bibinfo{title}{{Loss of a circadian adrenal
  corticosterone rhythm following suprachiasmatic lesions in the rat}}}.
\newblock {\emph{\JournalTitle{Brain Research}}} \textbf{\bibinfo{volume}{42}},
  \bibinfo{pages}{201--206}, \doiprefix\url{10.1016/0006-8993(72)90054-6}
  (\bibinfo{year}{1972}).

\bibitem{Stephan1972}
\bibinfo{author}{Stephan, F.~K.} \& \bibinfo{author}{Zucker, I.}
\newblock \bibinfo{journal}{\bibinfo{title}{{Circadian rhythms in drinking
  behavior and locomotor activity of rats are eliminated by hypothalamic
  lesions.}}}
\newblock {\emph{\JournalTitle{Proceedings of the National Academy of Sciences
  of the United States of America}}} \textbf{\bibinfo{volume}{69}},
  \bibinfo{pages}{1583--1586}, \doiprefix\url{10.1073/pnas.69.6.1583}
  (\bibinfo{year}{1972}).

\bibitem{Liu1997a}
\bibinfo{author}{Liu, C.}, \bibinfo{author}{Weaver, D.~R.},
  \bibinfo{author}{Strogatz, S.~H.} \& \bibinfo{author}{Reppert, S.~M.}
\newblock \bibinfo{journal}{\bibinfo{title}{{Cellular Construction of a
  Circadian Clock: Period Determination in the Suprachiasmatic Nuclei}}}.
\newblock {\emph{\JournalTitle{Cell}}} \textbf{\bibinfo{volume}{91}},
  \bibinfo{pages}{855--860}, \doiprefix\url{10.1016/S0092-8674(00)80473-0}
  (\bibinfo{year}{1997}).

\bibitem{Kawamura2008}
\bibinfo{author}{Kawamura, Y.}, \bibinfo{author}{Nakao, H.},
  \bibinfo{author}{Arai, K.}, \bibinfo{author}{Kori, H.} \&
  \bibinfo{author}{Kuramoto, Y.}
\newblock \bibinfo{journal}{\bibinfo{title}{{Collective Phase Sensitivity}}}.
\newblock {\emph{\JournalTitle{Physical Review Letters}}}
  \textbf{\bibinfo{volume}{101}}, \bibinfo{pages}{024101},
  \doiprefix\url{10.1103/PhysRevLett.101.024101} (\bibinfo{year}{2008}).

\bibitem{Ko2009}
\bibinfo{author}{Ko, T.-W.} \& \bibinfo{author}{Ermentrout, G.}
\newblock \bibinfo{journal}{\bibinfo{title}{{Phase-response curves of coupled
  oscillators}}}.
\newblock {\emph{\JournalTitle{Physical Review E}}}
  \textbf{\bibinfo{volume}{79}}, \bibinfo{pages}{016211},
  \doiprefix\url{10.1103/PhysRevE.79.016211} (\bibinfo{year}{2009}).

\bibitem{Kori2009}
\bibinfo{author}{Kori, H.}, \bibinfo{author}{Kawamura, Y.},
  \bibinfo{author}{Nakao, H.}, \bibinfo{author}{Arai, K.} \&
  \bibinfo{author}{Kuramoto, Y.}
\newblock \bibinfo{journal}{\bibinfo{title}{{Collective-phase description of
  coupled oscillators with general network structure}}}.
\newblock {\emph{\JournalTitle{Physical Review E}}}
  \textbf{\bibinfo{volume}{80}}, \bibinfo{pages}{036207},
  \doiprefix\url{10.1103/PhysRevE.80.036207} (\bibinfo{year}{2009}).

\bibitem{Levnajic2010}
\bibinfo{author}{Levnaji{\'{c}}, Z.} \& \bibinfo{author}{Pikovsky, A.}
\newblock \bibinfo{journal}{\bibinfo{title}{{Phase resetting of collective
  rhythm in ensembles of oscillators}}}.
\newblock {\emph{\JournalTitle{Physical Review E}}}
  \textbf{\bibinfo{volume}{82}}, \bibinfo{pages}{056202},
  \doiprefix\url{10.1103/PhysRevE.82.056202} (\bibinfo{year}{2010}).

\bibitem{Meijer2003}
\bibinfo{author}{Meijer, J.~H.} \& \bibinfo{author}{Schwartz, W.~J.}
\newblock \bibinfo{journal}{\bibinfo{title}{{In Search of the Pathways for
  Light-Induced Pacemaker Resetting in the Suprachiasmatic Nucleus}}}.
\newblock {\emph{\JournalTitle{Journal of Biological Rhythms}}}
  \textbf{\bibinfo{volume}{18}}, \bibinfo{pages}{235--249},
  \doiprefix\url{10.1177/0748730403253370} (\bibinfo{year}{2003}).

\bibitem{Nagano2003}
\bibinfo{author}{Nagano, M.} \emph{et~al.}
\newblock \bibinfo{journal}{\bibinfo{title}{{An abrupt shift in the day/night
  cycle causes desynchrony in the mammalian circadian center.}}}
\newblock {\emph{\JournalTitle{The Journal of neuroscience : the official
  journal of the Society for Neuroscience}}} \textbf{\bibinfo{volume}{23}},
  \bibinfo{pages}{6141--6151}, \doiprefix\url{23/14/6141 [pii]}
  (\bibinfo{year}{2003}).

\bibitem{Pittendrigh1991}
\bibinfo{author}{Pittendrigh, C.~S.}, \bibinfo{author}{Kyner, W.~T.} \&
  \bibinfo{author}{Takamura, T.}
\newblock \bibinfo{journal}{\bibinfo{title}{{The Amplitude of Circadian
  Oscillations: Temperature Dependence, Latitudinal Clines, and the
  Photoperiodic Time Measurement}}}.
\newblock {\emph{\JournalTitle{Journal of Biological Rhythms}}}
  \textbf{\bibinfo{volume}{6}}, \bibinfo{pages}{299--313}
  (\bibinfo{year}{1991}).

\bibitem{Abraham2010}
\bibinfo{author}{Abraham, U.} \emph{et~al.}
\newblock \bibinfo{journal}{\bibinfo{title}{{Coupling governs entrainment range
  of circadian clocks.}}}
\newblock {\emph{\JournalTitle{Molecular systems biology}}}
  \textbf{\bibinfo{volume}{6}}, \bibinfo{pages}{438},
  \doiprefix\url{10.1038/msb.2010.92} (\bibinfo{year}{2010}).

\bibitem{Coomans2015}
\bibinfo{author}{Coomans, C.~P.}, \bibinfo{author}{Ramkisoensing, A.} \&
  \bibinfo{author}{Meijer, J.~H.}
\newblock \bibinfo{journal}{\bibinfo{title}{{The suprachiasmatic nuclei as a
  seasonal clock}}}.
\newblock {\emph{\JournalTitle{Frontiers in Neuroendocrinology}}}
  \textbf{\bibinfo{volume}{37}}, \bibinfo{pages}{29--42},
  \doiprefix\url{10.1016/j.yfrne.2014.11.002} (\bibinfo{year}{2015}).

\bibitem{Meijer2007}
\bibinfo{author}{Meijer, J.~H.}, \bibinfo{author}{Michel, S.} \&
  \bibinfo{author}{Vansteensel, M.~J.}
\newblock \bibinfo{journal}{\bibinfo{title}{{Processing of daily and seasonal
  light information in the mammalian circadian clock}}}.
\newblock {\emph{\JournalTitle{General and Comparative Endocrinology}}}
  \textbf{\bibinfo{volume}{152}}, \bibinfo{pages}{159--164},
  \doiprefix\url{10.1016/j.ygcen.2007.01.018} (\bibinfo{year}{2007}).

\bibitem{Sumova2004}
\bibinfo{author}{Sumov{\'{a}}, A.}, \bibinfo{author}{Bendov{\'{a}}, Z.},
  \bibinfo{author}{Sl{\'{a}}dek, M.}, \bibinfo{author}{Kov{\'{a}}{\v{c}}ikova,
  Z.} \& \bibinfo{author}{Illnerov{\'{a}}, H.}
\newblock \bibinfo{journal}{\bibinfo{title}{{Seasonal Molecular Timekeeping
  Within the Rat Circadian Clock}}}.
\newblock {\emph{\JournalTitle{Physiological Research}}}
  \textbf{\bibinfo{volume}{53}}, \bibinfo{pages}{S167--S176}
  (\bibinfo{year}{2004}).

\bibitem{Pittendrigh1976f}
\bibinfo{author}{Pittendrigh, C.~S.} \& \bibinfo{author}{Daan, S.}
\newblock \bibinfo{journal}{\bibinfo{title}{{A functional analysis of circadian
  pacemakers in noctural rodents: I. The stability and lability of circadian
  frequency}}}.
\newblock {\emph{\JournalTitle{Journal of Comparative Physiology A}}}
  \textbf{\bibinfo{volume}{106}}, \bibinfo{pages}{223--252}
  (\bibinfo{year}{1976}).

\bibitem{VDL2009}
\bibinfo{author}{VanderLeest, H.~T.}, \bibinfo{author}{Rohling, J. H.~T.},
  \bibinfo{author}{Michel, S.} \& \bibinfo{author}{Meijer, J.~H.}
\newblock \bibinfo{journal}{\bibinfo{title}{{Phase shifting capacity of the
  circadian pacemaker determined by the SCN neuronal network organization.}}}
\newblock {\emph{\JournalTitle{PloS one}}} \textbf{\bibinfo{volume}{4}},
  \bibinfo{pages}{e4976}, \doiprefix\url{10.1371/journal.pone.0004976}
  (\bibinfo{year}{2009}).

\bibitem{Myung2012}
\bibinfo{author}{Myung, J.} \emph{et~al.}
\newblock \bibinfo{journal}{\bibinfo{title}{{Period coding of Bmal1 oscillators
  in the suprachiasmatic nucleus.}}}
\newblock {\emph{\JournalTitle{The Journal of neuroscience : the official
  journal of the Society for Neuroscience}}} \textbf{\bibinfo{volume}{32}},
  \bibinfo{pages}{8900--8918}, \doiprefix\url{10.1523/JNEUROSCI.5586-11.2012}
  (\bibinfo{year}{2012}).

\bibitem{Myung2015}
\bibinfo{author}{Myung, J.} \emph{et~al.}
\newblock \bibinfo{journal}{\bibinfo{title}{{GABA-mediated repulsive coupling
  between circadian clock neurons in the SCN encodes seasonal time}}}.
\newblock {\emph{\JournalTitle{Proceedings of the National Academy of Sciences
  of the United States of America}}} \textbf{\bibinfo{volume}{112}},
  \bibinfo{pages}{E3920--E3929}, \doiprefix\url{10.1073/pnas.1421200112}
  (\bibinfo{year}{2015}).

\bibitem{Evans2013}
\bibinfo{author}{Evans, J.~A.}, \bibinfo{author}{Leise, T.~L.},
  \bibinfo{author}{Castanon-Cervantes, O.} \& \bibinfo{author}{Davidson, A.~J.}
\newblock \bibinfo{journal}{\bibinfo{title}{{Dynamic Interactions Mediated by
  Nonredundant Signaling Mechanisms Couple Circadian Clock Neurons}}}.
\newblock {\emph{\JournalTitle{Neuron}}} \textbf{\bibinfo{volume}{80}},
  \bibinfo{pages}{973--983}, \doiprefix\url{10.1016/j.neuron.2013.08.022}
  (\bibinfo{year}{2013}).

\bibitem{Buijink2016}
\bibinfo{author}{Buijink, M.~R.} \emph{et~al.}
\newblock \bibinfo{journal}{\bibinfo{title}{{Evidence for Weakened
  Intercellular Coupling in the Mammalian Circadian Clock under Long
  Photoperiod}}}.
\newblock {\emph{\JournalTitle{Plos One}}} \textbf{\bibinfo{volume}{11}},
  \bibinfo{pages}{e0168954} (\bibinfo{year}{2016}).

\bibitem{Ramkisoensing2014}
\bibinfo{author}{Ramkisoensing, A.} \emph{et~al.}
\newblock \bibinfo{journal}{\bibinfo{title}{{Enhanced phase resetting in the
  synchronized suprachiasmatic nucleus network.}}}
\newblock {\emph{\JournalTitle{Journal of biological rhythms}}}
  \textbf{\bibinfo{volume}{29}}, \bibinfo{pages}{4--15},
  \doiprefix\url{10.1177/0748730413516750} (\bibinfo{year}{2014}).

\bibitem{Hannay2018}
\bibinfo{author}{Hannay, K.~M.}, \bibinfo{author}{Forger, D.~B.} \&
  \bibinfo{author}{Booth, V.}
\newblock \bibinfo{journal}{\bibinfo{title}{{Macroscopic Models for Networks of
  Coupled Biological Oscillators}}}.
\newblock {\emph{\JournalTitle{Science Advances}}}  (\bibinfo{year}{2018}).

\bibitem{Lu2016}
\bibinfo{author}{Lu, Z.} \emph{et~al.}
\newblock \bibinfo{journal}{\bibinfo{title}{{Resynchronization of circadian
  oscillators and the east-west asymmetry of jet-lag}}}.
\newblock {\emph{\JournalTitle{Chaos}}} \textbf{\bibinfo{volume}{26}},
  \bibinfo{pages}{094811}, \doiprefix\url{10.1063/1.4954275}
  (\bibinfo{year}{2016}).

\bibitem{Hannay2015}
\bibinfo{author}{Hannay, K.~M.}, \bibinfo{author}{Booth, V.} \&
  \bibinfo{author}{Forger, D.~B.}
\newblock \bibinfo{journal}{\bibinfo{title}{{Collective phase response curves
  for heterogeneous coupled oscillators}}}.
\newblock {\emph{\JournalTitle{Physical Review E}}}
  \textbf{\bibinfo{volume}{92}}, \bibinfo{pages}{022923},
  \doiprefix\url{10.1103/PhysRevE.92.022923} (\bibinfo{year}{2015}).

\bibitem{Foley2011}
\bibinfo{author}{Foley, N.~C.} \emph{et~al.}
\newblock \bibinfo{journal}{\bibinfo{title}{{Characterization of orderly
  spatiotemporal patterns of clock gene activation in mammalian suprachiasmatic
  nucleus.}}}
\newblock {\emph{\JournalTitle{The European journal of neuroscience}}}
  \textbf{\bibinfo{volume}{33}}, \bibinfo{pages}{1851--65},
  \doiprefix\url{10.1111/j.1460-9568.2011.07682.x} (\bibinfo{year}{2011}).

\bibitem{Lee2013}
\bibinfo{author}{Lee, J.~E.} \emph{et~al.}
\newblock \bibinfo{journal}{\bibinfo{title}{{Quantitative peptidomics for
  discovery of circadian-related peptides from the rat suprachiasmatic
  nucleus}}}.
\newblock {\emph{\JournalTitle{Journal of Proteome Research}}}
  \textbf{\bibinfo{volume}{12}}, \bibinfo{pages}{585--593},
  \doiprefix\url{10.1021/pr300605p} (\bibinfo{year}{2013}).

\bibitem{An2012}
\bibinfo{author}{An, S.}, \bibinfo{author}{Tsai, C.},
  \bibinfo{author}{Ronecker, J.}, \bibinfo{author}{Bayly, A.} \&
  \bibinfo{author}{Herzog, E.~D.}
\newblock \bibinfo{journal}{\bibinfo{title}{{Spatiotemporal distribution of
  vasoactive intestinal polypeptide receptor 2 in mouse suprachiasmatic
  nucleus}}}.
\newblock {\emph{\JournalTitle{The Journal of Comparative Neurology}}}
  \textbf{\bibinfo{volume}{520}}, \bibinfo{pages}{2730--2741},
  \doiprefix\url{10.1002/cne.23078} (\bibinfo{year}{2012}).

\bibitem{Maywood2006}
\bibinfo{author}{Maywood, E.~S.} \emph{et~al.}
\newblock \bibinfo{journal}{\bibinfo{title}{{Synchronization and maintenance of
  timekeeping in suprachiasmatic circadian clock cells by neuropeptidergic
  signaling}}}.
\newblock {\emph{\JournalTitle{Current Biology}}}
  \textbf{\bibinfo{volume}{16}}, \bibinfo{pages}{599--605},
  \doiprefix\url{10.1016/j.cub.2006.02.023} (\bibinfo{year}{2006}).

\bibitem{Farajnia2014a}
\bibinfo{author}{Farajnia, S.}, \bibinfo{author}{van Westering, T. L.~E.},
  \bibinfo{author}{Meijer, J.~H.} \& \bibinfo{author}{Michel, S.}
\newblock \bibinfo{journal}{\bibinfo{title}{{Seasonal induction of GABAergic
  excitation in the central mammalian clock.}}}
\newblock {\emph{\JournalTitle{Proceedings of the National Academy of Sciences
  of the United States of America}}} \textbf{\bibinfo{volume}{111}},
  \bibinfo{pages}{9627--9632}, \doiprefix\url{10.1073/pnas.1319820111}
  (\bibinfo{year}{2014}).

\bibitem{Dewoskin2015}
\bibinfo{author}{DeWoskin, D.} \emph{et~al.}
\newblock \bibinfo{journal}{\bibinfo{title}{{Distinct roles for GABA across
  multiple timescales in mammalian circadian timekeeping}}}.
\newblock {\emph{\JournalTitle{Proceedings of the National Academy of Sciences
  of the United States of America}}} \textbf{\bibinfo{volume}{112}},
  \bibinfo{pages}{E3911--E3919}, \doiprefix\url{10.1073/pnas.1420753112}
  (\bibinfo{year}{2015}).

\bibitem{Liu2000}
\bibinfo{author}{Liu, C.} \& \bibinfo{author}{Reppert, S.~M.}
\newblock \bibinfo{journal}{\bibinfo{title}{{GABA synchronizes clock cells
  within the suprachiasmatic circadian clock.}}}
\newblock {\emph{\JournalTitle{Neuron}}} \textbf{\bibinfo{volume}{25}},
  \bibinfo{pages}{123--128}, \doiprefix\url{10.1016/S0896-6273(00)80876-4}
  (\bibinfo{year}{2000}).

\bibitem{Freeman2013}
\bibinfo{author}{Freeman, G.~M.}, \bibinfo{author}{Krock, R.~M.},
  \bibinfo{author}{Aton, S.~J.}, \bibinfo{author}{Thaben, P.} \&
  \bibinfo{author}{Herzog, E.~D.}
\newblock \bibinfo{journal}{\bibinfo{title}{{GABA networks destabilize genetic
  oscillations in the circadian pacemaker}}}.
\newblock {\emph{\JournalTitle{Neuron}}} \textbf{\bibinfo{volume}{78}},
  \bibinfo{pages}{799--806}, \doiprefix\url{10.1016/j.neuron.2013.04.003}
  (\bibinfo{year}{2013}).

\bibitem{Aton2006}
\bibinfo{author}{Aton, S.~J.}, \bibinfo{author}{Huettner, J.~E.},
  \bibinfo{author}{Straume, M.} \& \bibinfo{author}{Herzog, E.~D.}
\newblock \bibinfo{journal}{\bibinfo{title}{{GABA and Gi/o differentially
  control circadian rhythms and synchrony in clock neurons.}}}
\newblock {\emph{\JournalTitle{Proceedings of the National Academy of Sciences
  of the United States of America}}} \textbf{\bibinfo{volume}{103}},
  \bibinfo{pages}{19188--19193}, \doiprefix\url{10.1073/pnas.0607466103}
  (\bibinfo{year}{2006}).

\bibitem{Mintz2002}
\bibinfo{author}{Mintz, E.~M.}, \bibinfo{author}{Jasnow, A.~M.},
  \bibinfo{author}{Gillespie, C.~F.}, \bibinfo{author}{Huhman, K.~L.} \&
  \bibinfo{author}{Albers, H.~E.}
\newblock \bibinfo{journal}{\bibinfo{title}{{GABA interacts with photic
  signaling in the suprachiasmatic nucleus to regulate circadian phase
  shifts}}}.
\newblock {\emph{\JournalTitle{Neuroscience}}} \textbf{\bibinfo{volume}{109}},
  \bibinfo{pages}{773--778}, \doiprefix\url{10.1016/S0306-4522(01)00519-X}
  (\bibinfo{year}{2002}).

\bibitem{Cao2013}
\bibinfo{author}{Cao, R.} \emph{et~al.}
\newblock \bibinfo{journal}{\bibinfo{title}{{Translational control of
  entrainment and synchrony of the suprachiasmatic circadian clock by
  mTOR/4E-BP1 signaling}}}.
\newblock {\emph{\JournalTitle{Neuron}}} \textbf{\bibinfo{volume}{79}},
  \bibinfo{pages}{712--724}, \doiprefix\url{10.1016/j.neuron.2013.06.026}
  (\bibinfo{year}{2013}).

\bibitem{Sumova1995}
\bibinfo{author}{Sumova, A.}, \bibinfo{author}{Travnickova, Z.},
  \bibinfo{author}{Peters, R.}, \bibinfo{author}{Schwartz, W.~J.} \&
  \bibinfo{author}{Illnerova, H.}
\newblock \bibinfo{journal}{\bibinfo{title}{{The rat suprachiasmatic nucleus is
  a clock for all seasons}}}.
\newblock {\emph{\JournalTitle{Proc Natl Acad Sci U S A}}}
  \textbf{\bibinfo{volume}{92}}, \bibinfo{pages}{7754--8},
  \doiprefix\url{10.1073/pnas.92.17.7754} (\bibinfo{year}{1995}).

\bibitem{Daido1996}
\bibinfo{author}{Daido, H.}
\newblock \bibinfo{journal}{\bibinfo{title}{{Onset of cooperative entrainment
  in limit-cycle oscillators with uniform all-to-all interactions: bifurcation
  of the order function}}}.
\newblock {\emph{\JournalTitle{Physica D}}} \textbf{\bibinfo{volume}{91}},
  \bibinfo{pages}{24--66} (\bibinfo{year}{1996}).

\bibitem{Daido1993}
\bibinfo{author}{Daido, H.}
\newblock \bibinfo{journal}{\bibinfo{title}{{Critical conditions of macroscopic
  mutual entrainment in uniformly coupled limit-cycle oscillators}}}.
\newblock {\emph{\JournalTitle{Progress of theoretical physics}}}
  \textbf{\bibinfo{volume}{89}}, \bibinfo{pages}{929--934}
  (\bibinfo{year}{1993}).

\bibitem{Ott2008}
\bibinfo{author}{Ott, E.} \& \bibinfo{author}{Antonsen, T.~M.}
\newblock \bibinfo{journal}{\bibinfo{title}{{Low dimensional behavior of large
  systems of globally coupled oscillators.}}}
\newblock {\emph{\JournalTitle{Chaos (Woodbury, N.Y.)}}}
  \textbf{\bibinfo{volume}{18}}, \bibinfo{pages}{037113},
  \doiprefix\url{10.1063/1.2930766} (\bibinfo{year}{2008}).

\bibitem{StHilaire2012}
\bibinfo{author}{St~Hilaire, M.~a.} \emph{et~al.}
\newblock \bibinfo{journal}{\bibinfo{title}{{Human phase response curve to a 1
  h pulse of bright white light}}}.
\newblock {\emph{\JournalTitle{The Journal of Physiology}}}
  \textbf{\bibinfo{volume}{590}}, \bibinfo{pages}{3035--3045},
  \doiprefix\url{10.1113/jphysiol.2012.227892} (\bibinfo{year}{2012}).

\bibitem{Pittendrigh1988a}
\bibinfo{author}{Pittendrigh, C.~S.}
\newblock \bibinfo{journal}{\bibinfo{title}{{The photoperiodic phenomena:
  seasonal modulation of the "day within".}}}
\newblock {\emph{\JournalTitle{Journal of biological rhythms}}}
  \textbf{\bibinfo{volume}{3}}, \bibinfo{pages}{173--188},
  \doiprefix\url{10.1177/074873048800300206} (\bibinfo{year}{1988}).

\bibitem{Brown2009}
\bibinfo{author}{Brown, T.} \& \bibinfo{author}{Piggins, H.}
\newblock \bibinfo{journal}{\bibinfo{title}{{Spatiotemporal Heterogeneity in
  the Electrical Activity of Suprachiasmatic Nuclei Neurons and their Response
  to Photoperiod}}}.
\newblock {\emph{\JournalTitle{Journal of Biological Rhythms}}}
  \textbf{\bibinfo{volume}{24}}, \bibinfo{pages}{44--54},
  \doiprefix\url{10.1177/0748730408327918} (\bibinfo{year}{2009}).

\bibitem{Yan2005}
\bibinfo{author}{Yan, L.}, \bibinfo{author}{Foley, N.~C.},
  \bibinfo{author}{Bobula, J.~M.}, \bibinfo{author}{Kriegsfeld, L.~J.} \&
  \bibinfo{author}{Silver, R.}
\newblock \bibinfo{journal}{\bibinfo{title}{{Two Antiphase Oscillations Occur
  in Each Suprachiasmatic Nucleus of Behaviorally Split Hamsters}}}.
\newblock {\emph{\JournalTitle{Journal of Neuroscience}}}
  \textbf{\bibinfo{volume}{25}}, \bibinfo{pages}{9017--9026},
  \doiprefix\url{10.1523/JNEUROSCI.2538-05.2005} (\bibinfo{year}{2005}).

\bibitem{Ohta2005}
\bibinfo{author}{Ohta, H.}, \bibinfo{author}{Yamazaki, S.} \&
  \bibinfo{author}{McMahon, D.~G.}
\newblock \bibinfo{journal}{\bibinfo{title}{{Constant light desynchronizes
  mammalian clock neurons}}}.
\newblock {\emph{\JournalTitle{Nature Neuroscience}}}
  \textbf{\bibinfo{volume}{8}}, \bibinfo{pages}{267--269},
  \doiprefix\url{10.1038/nn1395} (\bibinfo{year}{2005}).

\bibitem{Evans2012}
\bibinfo{author}{Evans, J.~A.}, \bibinfo{author}{Elliott, J.~A.} \&
  \bibinfo{author}{Gorman, M.~R.}
\newblock \bibinfo{journal}{\bibinfo{title}{{Individual Differences in
  Circadian Waveform of Siberian Hamsters under Multiple Lighting
  Conditions}}}.
\newblock {\emph{\JournalTitle{Journal of Biological Rhythms}}}
  \textbf{\bibinfo{volume}{27}}, \bibinfo{pages}{410--419},
  \doiprefix\url{10.1177/0748730412455915} (\bibinfo{year}{2012}).

\bibitem{Azzi2017a}
\bibinfo{author}{Azzi, A.} \emph{et~al.}
\newblock \bibinfo{journal}{\bibinfo{title}{{Network Dynamics Mediate Circadian
  Clock Plasticity}}}.
\newblock {\emph{\JournalTitle{Neuron}}} \textbf{\bibinfo{volume}{93}},
  \bibinfo{pages}{441--450}, \doiprefix\url{10.1016/j.neuron.2016.12.022}
  (\bibinfo{year}{2017}).

\bibitem{Farajnia2012}
\bibinfo{author}{Farajnia, S.} \emph{et~al.}
\newblock \bibinfo{journal}{\bibinfo{title}{{Evidence for neuronal desynchrony
  in the aged suprachiasmatic nucleus clock.}}}
\newblock {\emph{\JournalTitle{The Journal of neuroscience : the official
  journal of the Society for Neuroscience}}} \textbf{\bibinfo{volume}{32}},
  \bibinfo{pages}{5891--9}, \doiprefix\url{10.1523/JNEUROSCI.0469-12.2012}
  (\bibinfo{year}{2012}).

\bibitem{Nygard2005}
\bibinfo{author}{Nyg{\aa}rd, M.}, \bibinfo{author}{Hill, R.~H.},
  \bibinfo{author}{Wikstr{\"{o}}m, M.~A.} \& \bibinfo{author}{Kristensson, K.}
\newblock \bibinfo{journal}{\bibinfo{title}{{Age-related changes in
  electrophysiological properties of the mouse suprachiasmatic nucleus in
  vitro}}}.
\newblock {\emph{\JournalTitle{Brain Research Bulletin}}}
  \textbf{\bibinfo{volume}{65}}, \bibinfo{pages}{149--154},
  \doiprefix\url{10.1016/j.brainresbull.2004.12.006} (\bibinfo{year}{2005}).

\bibitem{Nygard2006}
\bibinfo{author}{Nyg{\aa}rd, M.} \& \bibinfo{author}{Palomba, M.}
\newblock \bibinfo{journal}{\bibinfo{title}{{The GABAergic network in the
  suprachiasmatic nucleus as a key regulator of the biological clock: does it
  change during senescence?}}}
\newblock {\emph{\JournalTitle{Chronobiology international}}}
  \textbf{\bibinfo{volume}{23}}, \bibinfo{pages}{427--35},
  \doiprefix\url{10.1080/07420520500545938} (\bibinfo{year}{2006}).

\bibitem{Benloucif1997}
\bibinfo{author}{Benloucif, S.}, \bibinfo{author}{Masana, M.~I.} \&
  \bibinfo{author}{Dubocovich, M.~L.}
\newblock \bibinfo{journal}{\bibinfo{title}{{Light-induced phase shifts of
  circadian activity rhythms and immediate early gene expression in the
  suprachiasmatic nucleus are attenuated in old C3H/HeN mice}}}.
\newblock {\emph{\JournalTitle{Brain Research}}}
  \textbf{\bibinfo{volume}{747}}, \bibinfo{pages}{34--42},
  \doiprefix\url{10.1016/S0006-8993(96)01182-1} (\bibinfo{year}{1997}).

\bibitem{Sellix2012}
\bibinfo{author}{Sellix, M.~T.} \emph{et~al.}
\newblock \bibinfo{journal}{\bibinfo{title}{{Aging Differentially Affects the
  Re-entrainment Response of Central and Peripheral Circadian Oscillators}}}.
\newblock {\emph{\JournalTitle{Journal of Neuroscience}}}
  \textbf{\bibinfo{volume}{32}}, \bibinfo{pages}{16193--16202},
  \doiprefix\url{10.1523/JNEUROSCI.3559-12.2012} (\bibinfo{year}{2012}).

\bibitem{Valentinuzzi1997}
\bibinfo{author}{Valentinuzzi, V.~S.}, \bibinfo{author}{Scarbrough, K.},
  \bibinfo{author}{Takahashi, J.~S.} \& \bibinfo{author}{Turek, F.~W.}
\newblock \bibinfo{journal}{\bibinfo{title}{{Effects of aging on the circadian
  rhythm of wheel-running activity in C57BL/6 mice}}}.
\newblock {\emph{\JournalTitle{The American journal of physiology}}}
  \textbf{\bibinfo{volume}{273}}, \bibinfo{pages}{1957--64}
  (\bibinfo{year}{1997}).

\bibitem{Biello2009}
\bibinfo{author}{Biello, S.~M.}
\newblock \bibinfo{journal}{\bibinfo{title}{{Circadian clock resetting in the
  mouse changes with age}}}.
\newblock {\emph{\JournalTitle{Age}}} \textbf{\bibinfo{volume}{31}},
  \bibinfo{pages}{293--303}, \doiprefix\url{10.1007/s11357-009-9102-7}
  (\bibinfo{year}{2009}).

\bibitem{Evans2016}
\bibinfo{author}{Evans, J.} \& \bibinfo{author}{Gorman, M.}
\newblock \bibinfo{journal}{\bibinfo{title}{{In synch but not in step:
  Circadian clock circuits regulating plasticity in daily rhythms}}}.
\newblock {\emph{\JournalTitle{Neuroscience}}} \textbf{\bibinfo{volume}{320}},
  \bibinfo{pages}{259--280}, \doiprefix\url{10.1016/j.neuroscience.2016.01.072}
  (\bibinfo{year}{2016}).

\bibitem{Abel2016}
\bibinfo{author}{Abel, J.~H.} \emph{et~al.}
\newblock \bibinfo{journal}{\bibinfo{title}{{Functional network inference of
  the suprachiasmatic nucleus}}}.
\newblock {\emph{\JournalTitle{Proceedings of the National Academy of
  Sciences}}} \textbf{\bibinfo{volume}{113}}, \bibinfo{pages}{4512--4517}
  (\bibinfo{year}{2016}).

\bibitem{Kori2006}
\bibinfo{author}{Kori, H.} \& \bibinfo{author}{Mikhailov, A.~S.}
\newblock \bibinfo{journal}{\bibinfo{title}{{Strong effects of network
  architecture in the entrainment of coupled oscillator systems}}}.
\newblock {\emph{\JournalTitle{Physical Review E}}}
  \textbf{\bibinfo{volume}{74}}, \bibinfo{pages}{066115},
  \doiprefix\url{10.1103/PhysRevE.74.066115} (\bibinfo{year}{2006}).

\bibitem{Kori2004}
\bibinfo{author}{Kori, H.} \& \bibinfo{author}{Mikhailov, A.~S.}
\newblock \bibinfo{journal}{\bibinfo{title}{{Entrainment of Randomly Coupled
  Oscillator Networks by a Pacemaker}}}.
\newblock {\emph{\JournalTitle{Physical Review Letters}}}
  \textbf{\bibinfo{volume}{93}}, \bibinfo{pages}{254101},
  \doiprefix\url{10.1103/PhysRevLett.93.254101} (\bibinfo{year}{2004}).

\end{thebibliography}




\section*{Acknowledgements (not compulsory)}
This research was supported by National Science Foundation grants DMS-1412119 and DMS-1853506. 

\section*{Author contributions statement}
 K.M.H. derived the results and wrote the simulations. V.B., D.B.F., and K.M.H. conceived the project, interpreted the results and collectively designed the study. All authors contributed to the writing of the manuscript. 

\section*{Additional information}

\subsection*{Competing Interests}
The authors declare no competing interests.



\end{document}